\documentstyle[12pt,aas2pp4]{article}
\begin{document}

\title{The Complex X-ray Spectra of M82 and NGC 253}

\author{A. Ptak\altaffilmark{1},
P. Serlemitsos,
T. Yaqoob\altaffilmark{2}, R. Mushotzky}
\affil{NASA/Goddard Space Flight Center, Code 662, Greenbelt, MD 20771, USA}
\and
\author{T. Tsuru}
\affil{Department of Physics, Faculty of Science, Kyoto University,
Sakyo-ku, Kyoto 606-01, Japan}
\altaffiltext{1}{University of Maryland, College Park, MD 20742, USA}
\altaffiltext{2}{With the Universities Space Research Association}
\received{Jul. 2, 1996}
\accepted{Dec. 20, 1996}
\slugcomment{To appear in the Apr. 1997 AJ}
\begin{abstract}
We present the results of the first imaging X-ray observations 
in the 0.4-10.0 keV bandpass of the nearby starburst galaxies 
M82 and NGC 253. The {\it ASCA} spectra of both M82 and NGC 253 are complex 
with 
strong line emission from O, Ne, Fe, Mg, S, and Si, allowing 
elemental abundances to be estimated in the X-ray band for the 
first time in these sources.  Two components are required to 
fit the spectra of both galaxies, with a ``soft'' component well 
described by a thermal model with a temperature of $\sim 10^{6-7}$ K and 
a ``hard'' component well described by either a thermal 
model ($T_{hard} \sim 10^{8}$ K) or a power-law model ($\Gamma 
\sim 1.8-2.0$). We find that different models (with different 
continua) yield absolute abundances that differ by more than an order of 
magnitude, while relative abundances are more robust and suggest 
an underabundance of Fe (inferred from the Fe-L complex) relative 
to $\alpha$-burning elements. Dust depletion may be responsible for the low
relative abundance of Fe inferred from the soft-component fits.
We set the most reliable limits to date 
on the Fe-K emission line equivalent width (EW), with an upper 
limit in NGC 253 of EW $<$ 180 eV while a line at $\sim$ 6.6 
keV is marginally significant (at the 3$\sigma$ level) with EW $\sim 100$ 
eV in M82. The low Fe-K line emission EW limits suggest either significantly 
sub-solar abundances in the material producing the hard component (if thermal) 
or that there is a 
significant non-thermal or non-equilibrium contribution to the 
hard component. Most of the soft flux (which originates mostly 
within the central kpc of M82 and NGC 253) is consistent with starburst 
models of supernovae-heated ISM and, to a lesser extent, starburst-driven 
superwind emission and the direct emission from supernova (SN). 
The hard component in both galaxies may have some contribution 
from $\sim 10^{8}$ K superwind emission or individual SN, although 
most of the emission probably originates in point sources (most 
likely blackhole candidates or mini-AGN with $M_{Edd} > 3-20 
M_{\odot}$) and, possibly, inverse-Compton scattering of IR photons. 
The similarity of the spectral X-ray characteristics 
of NGC 253 and M82 to some LINERs and low-luminosity AGN suggests 
a link between 
AGN and starbursts (e.g., both may contain an accretion-driven emission
component).
\end{abstract}

\section{Introduction} 
M82 and NGC 253 are two of the nearest ``starburst'' galaxies 
(we assume distances of 3.6 and 2.5 Mpc, respectively; \markcite{a11} Freedman 
{\it et al.} 1994; \markcite{a8}Davidge \& Prichett 1990) where extensive 
star formation 
in the $<$ 1 kpc-scale nuclear regions (\markcite{a22}Rieke 
{\it et al.} 1980) has resulted in a spectacular X-ray morphology, 
with emission concentrated both in the nucleus (at levels of 
one to two orders of magnitudes higher than typically seen in 
normal spirals) and extended along their minor axes to scales 
of up to $\sim$ 10 kpc. M82 and NGC 253 have been extensively 
studied in X-rays, with data from the 
{\it Einstein} HRI (\markcite{a37}Watson, Stanger, \& Griffiths 1984; 
\markcite{a10}Fabbiano 
\& Trinchieri 1984), 
{\it ROSAT} HRI  (\markcite{a2}Bregman, Schulman, \& Tomisaka 1995) and 
{\it EXOSAT} LE (\markcite{a23}Schaaf {\it et al.} 1989) revealing the X-ray 
morphology 
below $\sim$ 4 keV, showing that much of the nuclear emission 
is concentrated in multiple point sources. Data from the 
{\it Einstein} IPC (\markcite{a9}Fabbiano 1988), 
{\it EXOSAT} ME (\markcite{a23}Schaaf {\it et al.} 1989), 
{\it Ginga} (\markcite{a19}Ohashi {\it et al.} 1990; 
\markcite{a36}Tsuru {\it et al.} 1990) and 
{\it BBXRT} (\markcite{a21}Petre 1994) suggest a complex spectrum with at 
least two components, although the statistics, spectral resolution, 
and/or bandpass were somewhat limited with these instruments. 
These studies have suggested that the nuclear X-ray emission 
is due to an extended component probably consisting of the emission of 
unresolved
supernovae (SN) and SN-heated ISM, and individual
point sources which are likely to be very 
luminous X-ray binaries and young supernova (SN) or possibly even a hidden 
AGN.  These studies also suggest that the ``plumes'' breaking out of 
their galactic disks are evidence for SN-driven ``superwinds'' (Heckman,
Armus \& Miley 1990) 
based on the observed temperatures and luminosities. Inverse-Compton 
(IC) scattering of the large IR flux ($L_{IR} \sim 10^{10} L_{\odot}$; 
\markcite{a31}Telesco 1988) 
by relativistic electrons has also been suggested 
as a significant contributor to the X-ray flux (\markcite{a22}Rieke {\it et 
al.} 1980; \markcite{a23}Schaaf {\it et al.} 1989; \markcite{a17}Moran \& 
Lehnert 1996), a view (possibly) 
supported by the detection of MeV emission in the vicinity of 
NGC 253 (\markcite{a1}Bhattacharya {\it et al.} 1994). However, the expected 
intensity of the IC 
component is subject to large uncertainties depending on model 
assumptions (see \S 4). 

The significant spatial and spectral complexity suggests that 
proper deconvolution of the spectra would require good spatial 
{\it and} spectral resolution. 
Here we present high-quality 
spectra of M82 and NGC 253 in the 0.4-10.0 keV bandpass obtained 
by the X-ray satellite 
{\it ASCA} which carries the first imaging, medium-resolution 
X-ray spectrometers.
Although the spatial resolution of {\it ASCA} is
inferior to that of {\it ROSAT}, {\it ASCA} can constrain the source of
the
X-ray emission to within $\sim 1'$, while previous detectors with significant
effective area above 2 keV were essentially non-imaging (although
{\it BBXRT} allowed
for some crude spatial analysis).  We show below that the spectra are indeed
complex, showing strong line emission and requiring multiple components. 

\section{The ASCA Data} 
M82 was observed by 
{\it ASCA} (\markcite{a29}Tanaka, Inoue, \& Holt 1994) on May 19, 1993 and 
NGC 253 was observed on June 10, 1993. Briefly, 
{\it ASCA} consists of two solid-state imaging spectrometers 
(SIS; hereafter S0 and S1) with an approximate bandpass of 0.4-10.0 
keV and two gas imaging spectrometers (GIS; hereafter S2 and 
S3)  with an approximate bandpass of 0.8-10.0 keV. The SIS observations 
were done in 4-ccd mode although only the data from S0 chips 
1 and 2 and S1 chips 0 and 3 (containing most of the source counts) 
were used in the present analysis. Times of high background were 
excluded and hot pixels (in the SIS data) were removed. In M82 
and NGC 253, most of the extended emission is concentrated within 
$\sim$ 1' of the nucleus (\markcite{a10}Fabbiano \& Trinchieri 1984; 
\markcite{37}Watson, Stanger \& Griffiths 1984; 
\markcite{a2}Bregman, Schulman, \& Tomisaka 1995). 
 In NGC 253, several point sources are clearly present at distances 
of up to several arcminutes from the nucleus (see below).  Accordingly, 
source counts were extracted from a 6' radius circle centered on the 
nuclear source in each instrument (source regions sizes 
of 3' and 4' are typically chosen for the SIS and GIS, respectively, 
in the case of a single point source observation). The background 
was determined from the remaining counts in the chips beyond 
6.'5 for the SIS, from a 8-13' annulus for M82 GIS data and from 
an offset 5' circle for the NGC 253 GIS data. The observed M82 
count rates were 0.6-1.0 counts s$^{-1}$ (per detector)
among the four detectors with net exposure times of 
$\sim$ 17-20 ks and the background comprising 4-6\% of the total. 
The observed NGC 253 count rates were 0.1-0.3 counts s$^{-1}$
among the four detectors with net exposure times of 
$\sim$ 27-30 ks and the background comprising 12-19\% of the 
total.

{\it ROSAT} PSPC images of M82 and NGC 253 show that their extended halos
are soft (i.e., kT $\lesssim 0.5$ keV), contributing negligibly in the $\sim 
0.5-2.0$ 
keV bandpass 
compared to the X-ray flux originating within the 6' source
regions described above.  Nevertheless, it is possible that the background
regions used in extracting the {\it ASCA} background spectra may be
contaminated by the scattering of source photons or, possibly,
some
(relatively hard) non-thermal emission associated with the radio 
synchrotron halos
observed in the halos of M82 and NGC 253 (c.f., \markcite{a24}Seaquist \&
Odegard 1991; \markcite{a102} Carilli {\it et al.} 1992). 
The GIS spectra are particularly
vulnerable to these effects given the poorer spatial resolution and higher
sensitivity above 2 keV of the GIS relative to the SIS.  However, we find that
the counting rate in our background regions are consistent with spectra
extracted from blank sky fields.  Substituting the blank sky
background spectra for the background spectra described above does not
significantly affect any of the spectral results discussed below.
The telescope response was calculated assuming most of 
the counts within 6' are consistent with a point source (i.e., 
a source extent of $<$ 1').  
We went to some considerable effort
in investigating the effect of neglecting the spatial complexity
on the spectral analysis and these results are described in the
appendix. We find that  neglecting the spatial complexity
does not have a siginificant  impact on our results.
The spectra were binned to a minimum 
of 10 counts per bin to allow use of the $\chi^2$
statistic. 
%As shown below, our methods of data reduction 
%have resulted in consistent spectral fit results and fluxes among the four
%instruments. 

\section{Spectral Fitting} 
The SIS spectra strongly rule out a single-component model from 0.4-10.0 keV, 
with both power-law and Raymond-Smith 
plasma fits resulting in $\chi_{\nu}^{2} > 2.0$ for both sources. 
The poorer spectral resolution and response (at low energies) 
of the GIS allow for a single-component model. Indeed, most of 
the contribution to $\chi^{2}$
in the SIS fits results from narrow features below $\sim$ 
2.5 keV. This can be seen in Figure 1 which shows the ratio of data 
to best-fitting power-law model, with the location of expected 
K$_{\alpha}$ line emission from the He- and H-like ions of O, Ne, 
Mg, Si, S, Ca and Fe and the L-shell emission of Fe marked. The 
prominence of He-like Mg, Si and S lines implies a thermal component 
with kT $<$ 2 keV (c.f., \markcite{a16}Mewe, Gronenschild, \& Van Den Oord 
1985).
\begin{figure}
\plotone{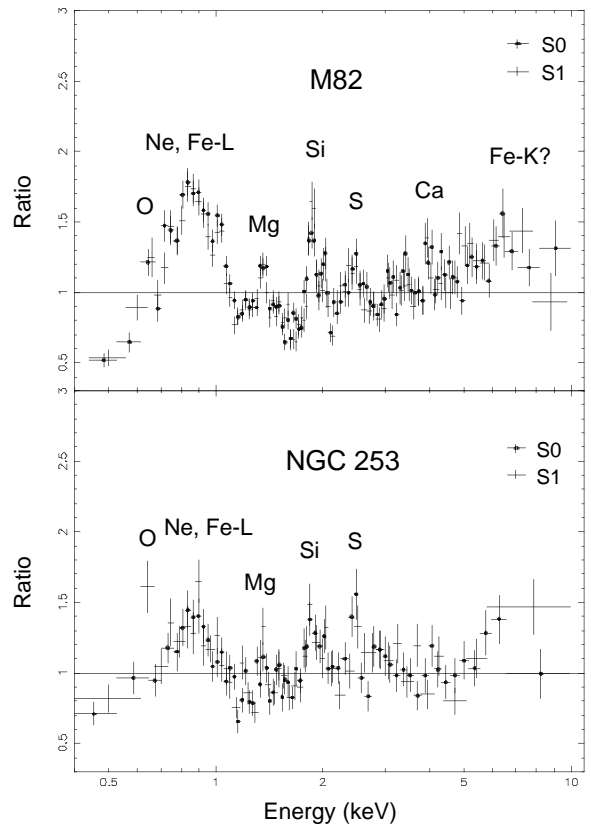}
\caption{Ratio of the SIS {\it ASCA} data to best-fitting power-law model 
for M82 (top)
and NGC 253 (bottom). The possible identification of prominent lines is
shown.}
\end{figure}
Accordingly we 
tried fitting two-component models to the spectra, with a Raymond-Smith 
model to account for the soft ($\lesssim$ 2 keV) flux and a Raymond-Smith 
or power-law component for the hard ($\gtrsim$ 2 keV) flux. 
These models resulted in acceptable fits for both NGC 253 and 
M82, although the SIS-only M82 fits resulted in $\chi^2_{\nu} \sim 1.3$ 
which can be rejected at a level of $>$99\% 
(note that this neglects any systematic error).
\renewcommand{\arraystretch}{1.2}
\begin{deluxetable}{cccccccc}
\tablewidth{0pt}
\tablenum{1}
%\tablewidth{108pt}
%\begin{center}
\tablecaption{Two-component Model Fits to {\it ASCA} M82 and NGC 253 Spectra }
%\begin{tabular}{cccccccc}
\tablehead{\colhead{Galaxy/Detector} & \colhead{Model$^*$} &
\colhead{$N_H^{soft}$} & \colhead{kT (keV)} & \colhead{$N_H^{hard}$} & 
\colhead{$\Gamma$/kT (keV)} & \colhead{$A/A_{\odot}^{\dagger}$} &
\colhead{$\chi^2$/dof}\nl}
\startdata
N253/S0-1 & 1 & $1.0^{+0.7}_{-1.0}$ & $0.81^{+0.06}_{-0.19}$ &
$15^{+15}_{-11}$ & $2.02^{+0.51}_{-0.29}$ & & 416.9/360 \nl % = 1.16\\
N253/S0-3 & 1 & $1.0^{+0.8}_{-1.0}$ & $0.80^{+0.06}_{-0.18}$ &
$14^{+13}_{-10}$ & $2.01^{+0.40}_{-0.40}$ & & 915.0/870 \nl %= 1.05 \nl
N253/S0-1 & 2 & $0.8^{+0.9}_{-0.8}$ & $0.81^{+0.06}_{-0.16}$ &
$8^{+13}_{-8}$ & $7.2^{+9.8}_{-3.2}$ & $0.03^{+0.32}_{-0.03}$ & 423.7/359 \nl
N253/S0-3 & 2 & $0.8^{+0.9}_{-0.8}$ & $0.81^{+0.06}_{-0.14}$ &
$8^{+11}_{-8}$ & $7.3^{+5.6}_{-2.7}$ & $0.02^{+0.23}_{-0.02}$ & 922.3/869 \nl
M82/S0-1$^{\ddag}$ & 1 & $2.4^{+0.7}_{-0.8}$ & $0.64^{+0.18}_{-0.32}$ &
$19^{+12}_{-19}$ & $1.78^{+0.35}_{-0.47}$ & & 548.1/422 \nl %= 1.30 \\
M82/S0-3 & 1 & $2.4^{+0.7}_{-0.6}$ & $0.63^{+0.13}_{-0.30}$ &
$19^{+10}_{-10}$ & $1.76^{+0.22}_{-0.21}$ & & 1379/1334 \nl  %= 1.03 \\
M82/S0-1 & 2 & $2.3^{+0.7}_{-2.3}$ & $0.63^{+0.15}_{-0.33}$ &
$14^{+11}_{-14}$ & $10^{+45}_{-4.}$ & $0.19^{+0.42}_{-0.19}$ &
540.7/421 \nl %= 1.28\\
M82/S0-3$^{\ddag}$ & 2 & $2.3^{+0.7}_{-0.8}$ & $0.63^{+0.13}_{-0.32}$ &
$14^{+10}_{-14}$ & $11.4^{+9.3}_{-4.4}$ & $0.18^{+0.35}_{-0.18}$ &
1366/1333 \nl %= 1.02\\
\enddata
%\hline \hline
%\end{tabular}
%\end{center}
\tablecomments{The errors give the 90\% confidence intervals assuming that all
model parameters (except for normalizations) are interesting.
Soft-component abundances derived from the 4-instrument Raymond-Smith plus 
power law fits are given in 
text. \nl
$^*$ 1: Raymond-Smith plus power law, 2: Raymond-Smith plus Raymond-Smith\nl
$^{\dagger}$ Hard-component abundance relative to solar \nl
$^{\ddag}$ Due to a relative minimum in $\chi^2$ space, the true soft 
component
absoption column lower confidence limit may be zero (see text).\nl 
$N_H^{soft}$ is in units of $10^{21} \ \rm cm^{-2}$, applied to both 
continuum components \nl
$N_H^{hard}$ is in units of $10^{21} \ \rm cm^{-2}$, applied to hard 
continuum component only \nl}
\end{deluxetable}

The results are 
shown in Table 1, where fits to the SIS data alone are denoted `S0-1' and fits
to the data from all four instruments are denoted `S0-3'. Table 1 shows that
the S0-1 and S0-3 fits give consistent results. The errors given in Table 1 
(and in the text below) assume that all model parameters are interesting except
for normalizations, or 12 (8) interesting parameters for the
M82 (NGC 253) Raymond-Smith plus power-law fits and 13 (9) interesting
parameters for the double Raymond-Smith fits (the M82 and NGC 253 fits have a
different number of model parameters due to a difference in the handling of the
soft-component abundances, as discussed below).
The double Raymond-Smith model gives a better 
fit to the M82 spectra while the Raymond-Smith plus power-law model 
gives a better fit to the NGC 253 spectra, although the differences 
in $\chi^{2}$ between the double Raymond-Smith and Raymond-Smith plus 
power-law models are statistically insignificant. The hard and 
soft components appear to be absorbed well in excess of the Galactic 
columns ($1.6 \times 10^{20} \ \rm cm^{-2}$ and $4.3 \times 10^{20} \ \rm 
cm^{-2}$ for NGC 253 and M82, respectively; \markcite{a26}Stark 
{\it et al.} 1992), although not at a high statistical significance. The
soft-component temperatures are similar ($\sim$ 
0.6 and 0.8 keV, for M82 and NGC 253, respectively, insensitive 
to the choice of hard-component continuum) while the hard component 
appears to be somewhat harder in M82 ($\Gamma \sim$ 1.8 or kT $\sim$ 11 keV) 
than in NGC 253 ($\Gamma \sim$ 2.0 or kT $\sim$ 7 keV). If the hard 
component is thermal, the abundance (inferred from a Raymond-Smith 
model) is constrained to be significantly sub-solar ($<$ 0.2 and 
$<$ 0.5 in NGC 253 and M82, respectively), based primarily on 
the lack of a strong Fe-K line at 6.7-6.9 keV. In NGC 253 no (narrow) Fe-K
line emission is detected, and we set a 90\% confidence upper-limit
(for two additional parameters, the
energy and normalization of the line) of $\sim$ 180 eV for the EW 
of any line emission in the four-instrument Raymond-Smith plus power-law fit.
In M82, the addition
of a narrow line at $\sim 6.6$ keV with an EW of $\sim 100$ eV to the
four-instrument Raymond-Smith plus power-law fit
reduces $\chi^2$ by $\sim 13$, significant
at a level of $\sim 3\sigma$ for two additional parameters. 
The Raymond-Smith plus power-law fits to the M82 and NGC 253 data 
are shown in Figure 2.
%an energy of $6.63^{+0.12}_{-0.22}$ keV and EW = $97^{+58}_{-60}$ eV, derived 
%from the four-instrument fit.
\begin{figure*}
%-ve offsets more the figure down and to the left
\plotfiddle{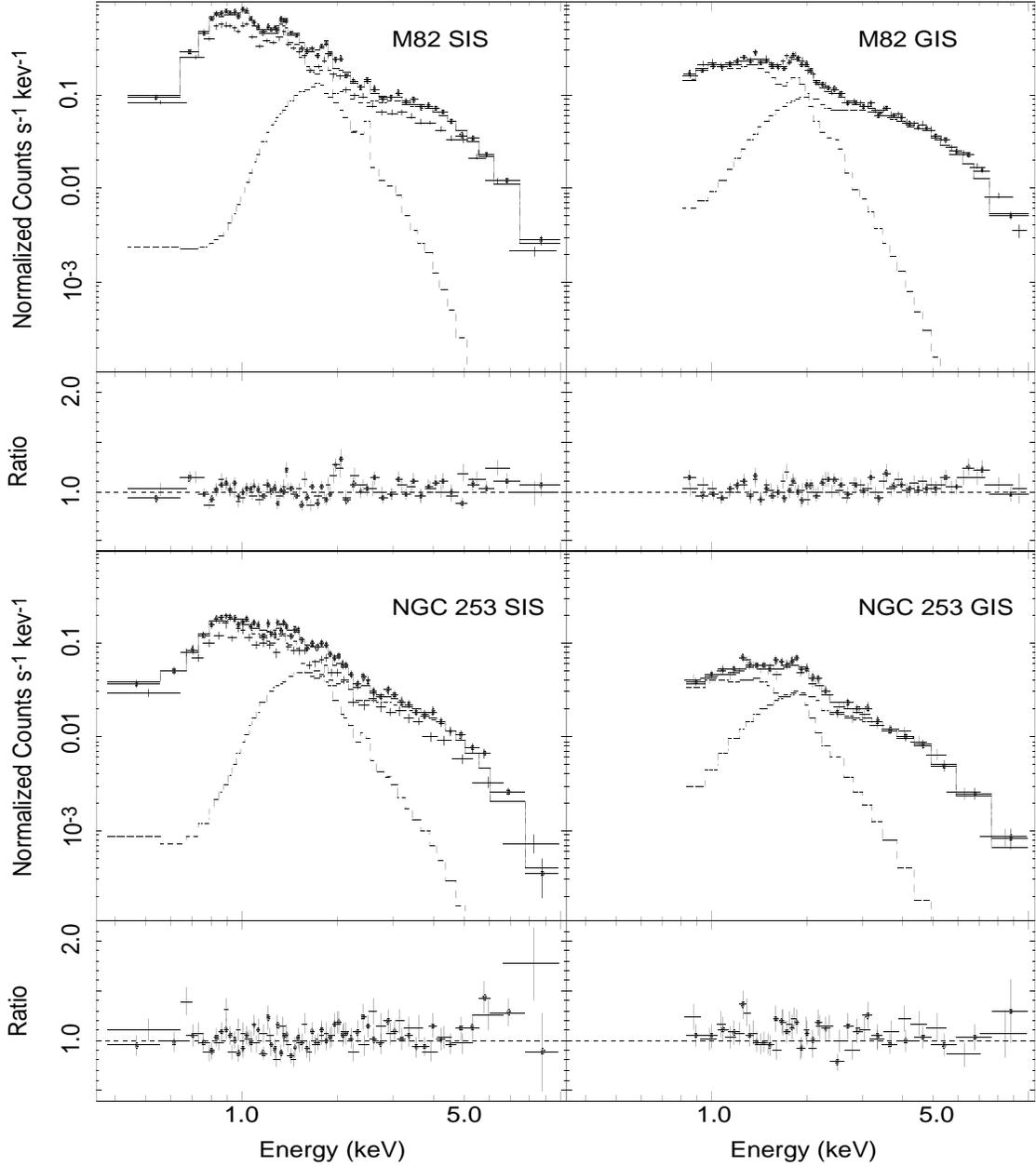}{6in}{0}{85}{75}{-262}{-70}
\caption{The best-fitting Raymond-Smith plus power-law model to the M82
and NGC 253 data.  The top panels show the data with the model plotted as a
solid line and the soft and hard components plotted separately as dashed 
lines, demonstrating that the hard component begins to dominate the spectra
of both M82 and NGC 253 above $\sim 2$ keV.
The SIS and GIS data are shown separately for clarity, with the S0 and S2
points marked.  The bottom panels show the ratio of the data to the models.
As is evident in these figures, the SIS have a broader bandpass, with a 
low-energy cutoff of $\sim 0.4$ keV, than the GIS, with a low-energy
cutoff of $\sim 0.8$ keV, while the GIS have more response at higher energies.}
\end{figure*}

In M82 the soft-component abundances (for the S0-3 Raymond-Smith 
plus power-law fit) relative to solar were found to be: N $<2.1$, 
O = $0.18^{+0.25}_{-0.14}$, Ne = $0.15^{+0.20}_{-0.14}$, Mg = $0.22^{+0.14}_
{-0.11}$, Si = $0.35^{+0.17}_{-0.11}$, S = $0.60^{+0.57}_{-0.37}$, Fe = 
$0.04^{+0.02}_{-0.03}$. 
The Ar, Ca, and Ni abundances were tied together (due to limited
signal-to-noise) 
resulting in a mean value of $0.23^{+0.36}_{-0.23}$. Note that the 
Fe abundance inferred from the soft-component model is determined 
solely based on Fe-L emission, not Fe-K, due to the dominance 
of the hard component above $\sim$ 2 keV. We note that these 
abundances were derived assuming that the soft component is 
{\it entirely thermal}. 
The poorer signal-to-noise of the NGC 253 
data do not warrant allowing the elemental abundances to vary 
independently. O, Mg, Si, and S, which show strong line emission 
in Figure 1, were grouped together, and Ne and Fe-L, which are 
not resolved from each other, were grouped separately. The resulting 
abundances were  N, Ar, Ca, Ni $<$ 0.66; O, Mg, Si, S = $0.27^{+0.67}_{-0.14}$ 
and Ne, Fe = $0.05^{+0.15}_{-0.04}$. As in M82, these results 
are suggestive of a depression of Fe abundance relative to O, 
Mg, Si and S.

The somewhat high value of $\chi^2$ resulting from the SIS-only
fits to the M82 data (see Table 1) suggests that a third component may be
required.  Relative minima in $\chi^2$ space are present in the M82 fits when
the soft-component 
temperature and $N_H$ are varied.  This results in a highly assymmetric
confidence interval for the temperature (see Table 1).  In the case of $N_H$,
the required difference in $\chi^2$ is reached before the onset of the new
minima, giving the lower limit shown in Table 1, however the ``true'' lower
limit on $N_H$ may be zero.  This
behavior also suggests the presence of an additional component, with a lower
temperature than that of the soft component.  The addition of a third thermal
component to the 
double Raymond-Smith model substantially improves the M82 fits, resulting in 
temperatures of 
0.35 keV, 1.07 keV and 12.4 keV (see \markcite{a35} Tsuru {\it et al.}, 1997
for a discussion of three-component Raymond-Smith and non-equilibrium plasma
fits to the M82 spectra).  Adding an additional thermal component to the
double Raymond-Smith NGC 253 model reduced $\chi^{2}$ by only 8.3 for
2 additional 
degrees of freedom (kT and normalization, with $N_{H}$ fixed 
at the Galactic value and the abundances tied to the T $\sim 10^{7}$ K 
component abundances), significant only at the $2.4\sigma$ 
level. The temperature of the additional component was $\sim$ 
0.14 keV and the best-fitting model parameters of the other two components
were similar to the two-component fit values in Table 1. 

\begin{figure}
\plotfiddle{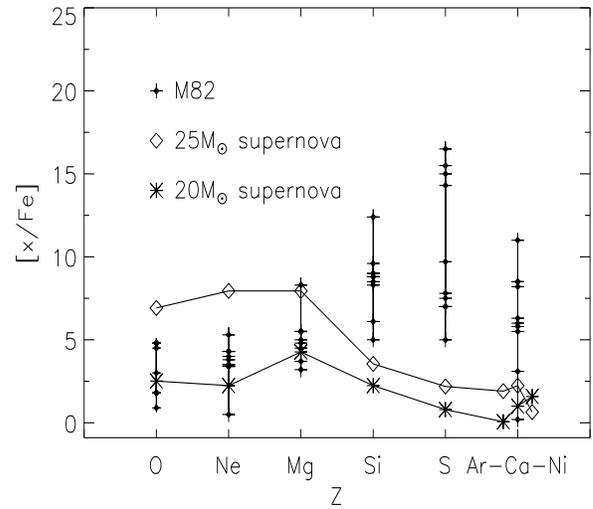}{2.0in}{90}{50}{60}{195}{-77}
\caption{The abundances of various elements relative to Fe. The multiple 
points plotted
for each element show the range of {\it relative} abundances derived from
various
models (see text). Note that the ranges shown are dominated by systematic
errors which are much larger than the statistical errors ({\it not}
explicitly shown), such as those for
the fits given in Table 1.  Also shown are the yields predicted for
$20M_{\odot}$ and $25M_{\odot}$ supernovae, taken from %\markcite{a32} 
Thielemann, Nomoto, \& 
Hashimoto (1996), normalized to Fe.}
\end{figure}
In Figure 3 we plot the abundances relative to Fe (x/Fe) derived 
from various fits to the M82 data.  
In addition to the M82 fits in Table 1 and the triple Raymond-Smith model
discussed above, these fits
include substituting the Mewe-Kaastra-Liedahl plasma model (`mekal' in XSPEC)
for the Raymond-Smith
plasma model in the Raymond-Smith plus power law model, a power-law
distribution of temperatures model, a double Raymond-Smith plus power-law 
model, 
and including an additional (very steep) 
power-law in the triple Raymond-Smith model. 
While these fits result in abundances that are {\it one to two orders of 
magnitude supersolar} (indicating that the soft-component continuum is poorly
constrained), Figure 3 shows that the {\it relative} abundances are 
somewhat more robust, with most varying by less than a factor of $\sim$ 5, 
although
the relative abundances of Ar, Ca and Ni are poorly constrained.  Note that
these errors are still somewhat larger than the statistical errors discussed
above, showing that systematic errors dominate. These systematic errors are
most likley dominated by differences in the slope of the continuum across
the line-emitting energies in each fit.  A comparison of the Raymond-Smith
and the Mewe-Kaastra-Liedahl models in XSPEC for a 0.6 keV, solar-abundance
plasma shows that, with
the exeption of an Oxygen line at $\sim 0.8$ keV present in the Raymond-Smith
model but not in the mekal model, most of the
differences between the two models are on the order of 30\% or less, implying
that uncertainties in atomic physics are likely to be much less significant
than the statistical errors in these data.
Also shown in Figure 3 are the x/Fe ratios expected for 20 
and 25 $M_{\odot}$ supernova showing that even the direct emission 
of Type-II supernovae should not show such a large (relative) 
underabundance of Fe, particularly relative to Si and S (\markcite{a32}
Thielemann {\it et al.} 1996).

\begin{table*}
\tablenum{2}
\caption{{\it ASCA} Fluxes and Luminosities for M82 and NGC 253}
\begin{center}
\begin{tabular}{|c|c|c|c|c|c|c|c|}
\hline
Galaxy & Energy Range & Observed Flux & 
\multicolumn{5}{|c|}{Absorption-Corrected Luminosities} \\
& & ($10^{-12} \ \rm ergs \ s^{-1} \ cm^{-2}$) &
\multicolumn{5}{|c|}{($10^{40} \ \rm ergs \ s^{-1}$)} \\
\cline{4-8}
& & & \multicolumn{2}{|c|}{Two-comp. Fit$^\dagger$} &
\multicolumn{3}{|c|}{Three-comp. Fit$^*$} \\
\cline{4-8}
& & & Soft & Hard & Soft & Medium & Hard  \\
\hline
M82 & 0.5-2.0 keV & 9.4 & 2.7 & 2.1 & 3.2 & 2.5 & 1.4 \\
M82 & 2.0-10.0 keV & 20 & 0.2 & 3.5 & 0.0 & 0.6 & 3.4 \\
NGC 253 & 0.5-2.0 keV & 2.8 & 0.23 & 0.33 & 0.02 & 0.26 & 0.19 \\
NGC 253 & 2.0-10.0 keV & 4.7 & 0.03 & 0.37 & 0.00 & 0.03 & 0.35 \\
\hline
\end{tabular}
\end{center}
All values are averages from S0-3 fits [M82 (NGC 253) fluxes from
individual detectors are within 3\% (4\%) of the mean]. \\
Observed fluxes are inferred from the best-fitting two-component model 
(three-component models give
similar results). \\
$^\dagger$ Power law (hard) plus Raymond-Smith (soft) model\\
$^*$ Triple Raymond-Smith model \\
\end{table*}

In Table 2 we give the observed fluxes and absorption-corrected 
luminosities (given for each component separately) for NGC 253 
and M82 in the 0.5-2.0 keV and 2-10 keV bands, with the SIS values 
corrected for non-uniform spatial exposure. The fluxes inferred 
from the individual detectors are within 3-4\% of the mean values, 
although systematic differences of 10-20\% between the SIS and 
GIS values are often found in other sources. The uncertainty in N$_{H}$ 
is significant in computing the unabsorbed luminosities, with 
the 90\% confidence interval in $N_{H}^{soft}$ in NGC 253 corresponding 
to an uncertainty of $\sim$ 64\% in intrinsic 
0.5-2.0 keV luminosity. The 0.5-2.0 keV PSPC fluxes for M82 and 
NGC 253 from 6' regions are $\sim$ 30\% and 10\% higher than 
the {\it ASCA} fluxes, respectively, and lie within the scatter of 
the correlation of the {\it ASCA} and 
{\it PSPC} fluxes for 13 low-activity galaxies (including M82 
and NGC 253) shown in \markcite{a25}Serlemitsos, Ptak \& Yaqoob (1996). Our 
2-10 keV fluxes are $\sim$ 1/2 the Ginga values (\markcite{a19}Ohashi 
{\it et al.} 1990, \markcite{a36}Tsuru {\it et al.} 1990) although any 
interpretation of this result 
must be viewed with caution given the large (1\arcdeg \ by 2\arcdeg) 
{\it Ginga} FOV (however, the fact that the M82 observation was performed in a
scanning mode reduces the uncertainty in flux).

\section{Discussion}
\subsection{Comparison with Superwind Models} 
While two and three-component models describe the data well, the 
true situation, particularly for the soft component below $\sim$ 
2 keV, is likely to be a distribution of temperatures. For example, 
the ROSAT PSPC data from both M82 and NGC 253 suggest temperatures 
of $\sim$ 10$^{7}$ K in their disk and nuclear regions, with 
the temperature decreasing with radius out in the halos (i.e., 
beyond $\sim$ 1 kpc), as discussed in \markcite{a28}Strickland, Ponman, \& 
Stevens (1996) and \markcite{a7}Dahlem, Heckman \& Weaver (1996). 
In NGC 253, the 
soft ($<$ 2 keV) component is apparently dominated by a disk 
and nuclear thermal component with kT $\sim$ 0.8 keV.  In M82, 
on the other hand, the soft component appears to have approximately 
equal contributions from $\sim$ 1.1 keV gas in the nuclear and 
disk regions and $\sim$ 0.4 keV gas in the halo, which is fit reasonably well
by a 0.8 keV thermal model. The luminosities, 
pressures, masses and energetics implied by the best-fitting 
soft-component model parameters are consistent with recent models 
of starburst-driven winds (\markcite{a33}Tomisaka \& Bregman 1993; 
\markcite{28}Suchkov 
{\it et al.} 1996), although the details of this kind of analysis 
are sensitive to assumptions concerning the volume, filling factor, 
and possibly the outflow velocity of the gas associated with 
each component.  For example, the outflow velocity of the X-ray 
emitting gas in M82 may be as high as 1000-3000 km s$^{-1}$ 
(\markcite{a24}Seaquist \& Odegard 1991) although the outflow velocity of 
molecular 
gas in M82 is estimated to be less than 500 km s$^{-1}$ (\markcite{a18}Nakai 
{\it et al.} 1987).  
To estimate the contribution of SN to the observed thermal flux we assume 
that the ISM heating is dominated by type-II SN, and that each supernova 
ejects $\sim 10 M_{\odot}$ with $10^{51}$ 
ergs of kinetic energy (c.f., \markcite{a28}Suchkov {\it et al.} 1996). 
With these assumptions, {\it independent of the assumed volume V and filling 
factor},
\begin{eqnarray*}
\frac{M_{SN}}{M_{total}} & \sim &
\frac{(\frac{3}{2}nVkT + \frac{1}{2}m_HnVv_{1000}^2)M_{10}}
{10^{51}m_HnV} \\
& \sim & \frac{\frac{3}{2} (kT+m_Hv_{1000}^2)M_{10}} {10^{51}m_H}
\end{eqnarray*}
where 
$M_{SN}$ is
the mass of gas ejected by SN (i.e., individual SN and the superwind),
$M_{total}$ is the total X-ray emitting
mass, $M_{10}$ is the mass ejected by a typical SN in units of $10M_{\odot}$
and $v_{1000}$ is the outflow velocity in units of $1000 \rm \ km \ s^{-1}$. 
We estimate 
(with $M_{10}$ = 1 and $v_{1000}$ = 1) that $\frac{M_{SN}}{M_{total}} \sim
12-14\%$ for the 1.1 keV and 0.8 keV gas
(for M82 and NGC 253, respectively) of the soft-component 
emission.  $M_{10}$ and $v_{1000}$ are likely to be somewhat smaller than 
$\sim 3$, 
so this implies that a significant amount of the emission 
is due to ambient ISM, possibly ``swept-up'' by the superwind, as 
suggested by \markcite{a28}Suchkov {\it et al.} (1996).
The low absolute abundances implied by some 
of the models of the soft component may therefore be indicative 
of low abundances in the (ambient and swept-up) disk ISM. This calculation 
is not 
appropriate for the $\sim$ 0.4 keV halo gas in M82 since energy 
dissipation is likely to significant out in the halo.  The unusually 
high values of x/Fe shown in Figure 3 indicate that the Fe abundance 
is ``suppressed'', particularly relative to S and 
Si.  The high IR luminosities of M82 and NGC 253 suggest a 
significant dust distribution in the disk and halo ISM of M82 and 
NGC 253 (\markcite{a31}Telesco 1988). 
Since Fe is depleted in warm ISM clouds to a larger extent than S and Si (c.f.,
Cowie \& Songaila 1986), dust depletion may account for some of the Fe 
deficiency, resulting in Si/Fe and S/Fe ratios $\sim$ 17\% and 130\%
(respectively) {\it higher} than in the undepleted case
\footnote{The dust depletion given in table 3 of \markcite{a6} Cowie \& 
Songalia was 
scaled to the $\sim$ 50\% depletion of Si
estimated by \markcite{a15}Lord {\it et al.} (1996) for the nuclear HII 
regions in M82 in this calculation.}. 

Repeating 
the ``swept-up mass'' calculation described above for the hard 
component, we find that, if thermal, $\sim$ 25-75\% (NGC 253) 
and 40-100\% (M82) of the hard component originates in superwind 
emission and 
{\it not} the ambient ISM. The high absorption ($N_H \sim 10^{22} \rm \ 
cm^{-2}$) derived from fits to the hard components (see Table 1) also suggests
that the hard component originates in compact and/or nuclear regions
(note that radio observations of the nuclear regions of both M82 and NGC 253 
suggest the presence of molecular gas with similar column densities; 
\markcite{a18}Nakai {\it et al.} 1986; \markcite{a20}Paglione, Tosaki, \& 
Jackson 1996).
However, the inconsistency of the low abundances
inferred from the hard-component fits with the high metallicity expected 
in supernova ejecta and solar abundances expected in the nuclear 
HII regions of M82 and NGC 253 (c.f., \markcite{a4}Carral {\it et al.} 1993; 
\markcite{a15}Lord 
{\it et al.} 1996) suggest that the hard components may not be dominated 
by (nuclear) superwind emission. Dust depletion may again be important 
although dust destruction is likely to be significant in nuclear starburst
regions over the $\sim 10^{7-8}$ year lifetime of the starburst (\markcite{a3}
Calzetti, Kinney \& Storchi-Bergmann 1996).  
It is therefore likely that gas not in 
thermal equilibrium or non-thermal emission contribute significantly 
to the hard components. The former case would correspond to $n_{0} 
< 0.5-1.5 \times 10^{-3} (v_{1000}/r_{kpc}) \rm \ cm^{-3}$, 
where $r_{kpc}$ is the hard-component emitting region size in 
kpc, $v_{1000}$ is the velocity of the ejecta (as defined above), and 
$n_{0}$ is the pre-shock ambient density 
(c.f., Tomisaka \& Bregman 1993; Hamilton, Sarazin, \& Chevalier 1983). 
Even if $v_{1000}$ is as high 
as 3, such a low value of $n_{0}$ is unlikely in the nuclear 
regions of M82 and NGC 253 (\markcite{a4}Carral 
{\it et al.} 1993; \markcite{a15}Lord 
{\it et al.} 1996), suggesting that non-thermal equilibrium ionization 
is not important unless the direct emission of SN is significant. 
Using the radio surface brightness - radius relationship for 
SN in \markcite{a14}Huang {\it et al.} (1993) and the X-ray luminosities of 
SN1986J and 
Cas A as a guide, we expect the X-ray luminosity of SN in M82 
to be $\sim 10^{37-38} \rm \ ergs \ s^{- 1}$, too low to account 
for more than a few percent of the hard component, while in NGC 
253 some of the bright extranuclear sources may be supernovae 
contributing to the hard component. Future work will examine 
the possible thermal contribution to the hard component in more 
detail, but it appears likely that the hard components in M82 
and NGC 253 are dominated by non-thermal emission.

\begin{figure}
\plotfiddle{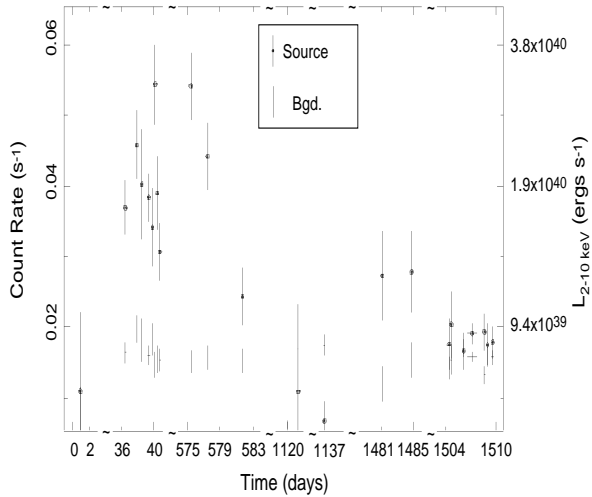}{2in}{0}{45}{60}{-130}{-210}
%\plotone{m82_n253_paper_fig4.eps}
\caption{The {\it ROSAT} HRI background-subtracted light curve for the
nuclear source in M82 (c.f., Figure 2 of
Collura {\it et al.} 1994), 
including recent AO-5 data. 
The source counts were extracted from a 10'' region centered on the source and
the background was determined from a 15''-20'' annulus centered on the source
(here the background is completely dominated by nuclear emission in M82).
Also shown is the corresponding absorption-corrected 2-10 keV luminosity
assuming a spectrum similar to the hard component inferred from the {\it ASCA} 
data (i.e., $\Gamma \sim 1.8$ and $N_H \sim 10^{22} \ \rm cm^{-2}$). 
Evidently, this source could account for a considerable fraction of the
intrinsic 2-10 keV luminosity of M82 ($L_X \sim 3.4 \times 10^{40} \rm \
ergs \ s^{-1}$) observed by {\it ASCA} (on day 755 in this figure).}
\end{figure}

\subsection{The Nature of the Non-thermal Emission}

The two most likely candidates for non-thermal emission are accretion-driven 
point sources (e.g., X-ray binaries and AGN) and inverse-Compton 
(IC) scattering of (nuclear) IR flux by relativistic electrons. 
The IC component may be significant (estimated by Schaaf {\it et al.} 
1991 to be $L_{1.4-8.9 \rm \ keV} \sim 1.5 \times 10^{40} \rm \ ergs \ s^{-1}$
or $\sim$ 50\% of the hard component in M82), although the 
likely high ratio of proton to electron energies (i.e., $\sim$ 
100; \markcite{a24}Seaquist \& Odegard 1991) would significantly reduce the 
contribution of this component. On the other hand, the point 
sources in M82 and NGC 253 may account for much of the hard component. 
In M82, a nuclear source has varied from $\sim$ 0.04-0.06 counts 
s$^{-1}$ to $\sim$ 0.02 counts s$^{- 1}$ between the 1991 (\markcite{a5}
Collura {\it et al.} 1994) and 
1995 ROSAT HRI observations, corresponding to an intrinsic 2-10 
keV luminosity of $1-4 \times 10^{40} \rm \ ergs \ s^{-1}$ or 
$\sim$ 30\%-100\% 
of the hard-component luminosity assuming this source has a spectrum 
similar to the power-law model fit to the hard component (see Table 
1 and Figure 4). In NGC 253, four strong extranuclear point sources ($\sim$ 
2-4' from the nucleus) are responsible for $\sim$ 20\% of the 
0.5-2.0 keV flux in the ROSAT PSPC observation. As shown in Figure 
5 (plate 1), the point sources in NGC 253 have varied between 
the {\it Einstein} and {\it ROSAT} observations and are clearly prominent 
above 2 keV in the 
{\it ASCA} S0 image. We extracted composite SIS spectra for the two brightest 
extranuclear sources (sources R1 and R5 in figure 5), using source
region sizes of 45'' to minimize the ``contamination'' from diffuse gas.  
Fitting
these spectra with the Raymond-Smith plus power-law model (see Table 1) 
resulted in fit parameters consistent with those listed in Table 1 for 
the entire galaxy (e.g., $N_H \sim 2.1^{+7.4}_{-1.9} \times 10^{22} \ \rm
cm^{-2}$ and $\Gamma \sim 2.3^{+1.7}_{-1.0}$).  These fits suggest that sources
R1 and 5 account for $\sim 25\%$ of the observed 2-10 keV flux from NGC 253.
While it is possible that some of the
point sources in NGC 253 are associated with SN (see above), some are likely 
to be 
X-ray binaries given the lack of optical counterparts (i.e., 
\markcite{a10}Fabbiano \& Trinchieri 1984). The Eddington luminosities of 
these sources suggest masses of at least 3-20 $M_{\odot}$ implying that 
they are blackhole candidates, which also have spectra consistent 
with the hard components observed in M82 and NGC 253 (Tanaka 
1989). The spectra of M82 and NGC 253 are also very similar to 
the spectra of low-luminosity AGN and LINERs (Serlemitsos, Ptak 
and Yaqoob 1996) which generally can be described by a Raymond-Smith 
plus power-law model. The power-law slopes obtained from our fits 
($\Gamma \sim 1.8-2.2$) may suggest a common physical origin, such 
as an accretion-driven component. It is likely that some of the 
point sources in M82 and NGC 253 have masses $\gg 10 M_{\odot}$ 
since it would be unphysical for all of these sources to be emitting 
at their Eddington luminosities most of the time, if they are 
powered by accretion.
%The (possible) long-term variability suggested 
%by comparing the {\it ASCA} and 
%{\it Ginga} observations of M82 and NGC 253 would be similar 
%to variability observed in LINERs and LLAGN (c.f., M81).
The discovery of (long-term) X-ray variability in ``starburst'' galaxy NGC 3628
demonstrates that the X-ray flux from NGC 3628 is dominated by a point
source, most likely a massive blackhole candidate or a micro-AGN 
(\markcite{a100} Dahlem, Heckman, Fabbiano 1995; \markcite{a101} 
Yaqoob {\it et al.} 1995).
These considerations suggest the possibility that starburst galaxies such as
M82 and NGC 253 may harbor blackhole 
candidates, with masses significantly higher than the masses of blackhole
candidates in our galaxy.

\begin{figure*}
\plotfiddle{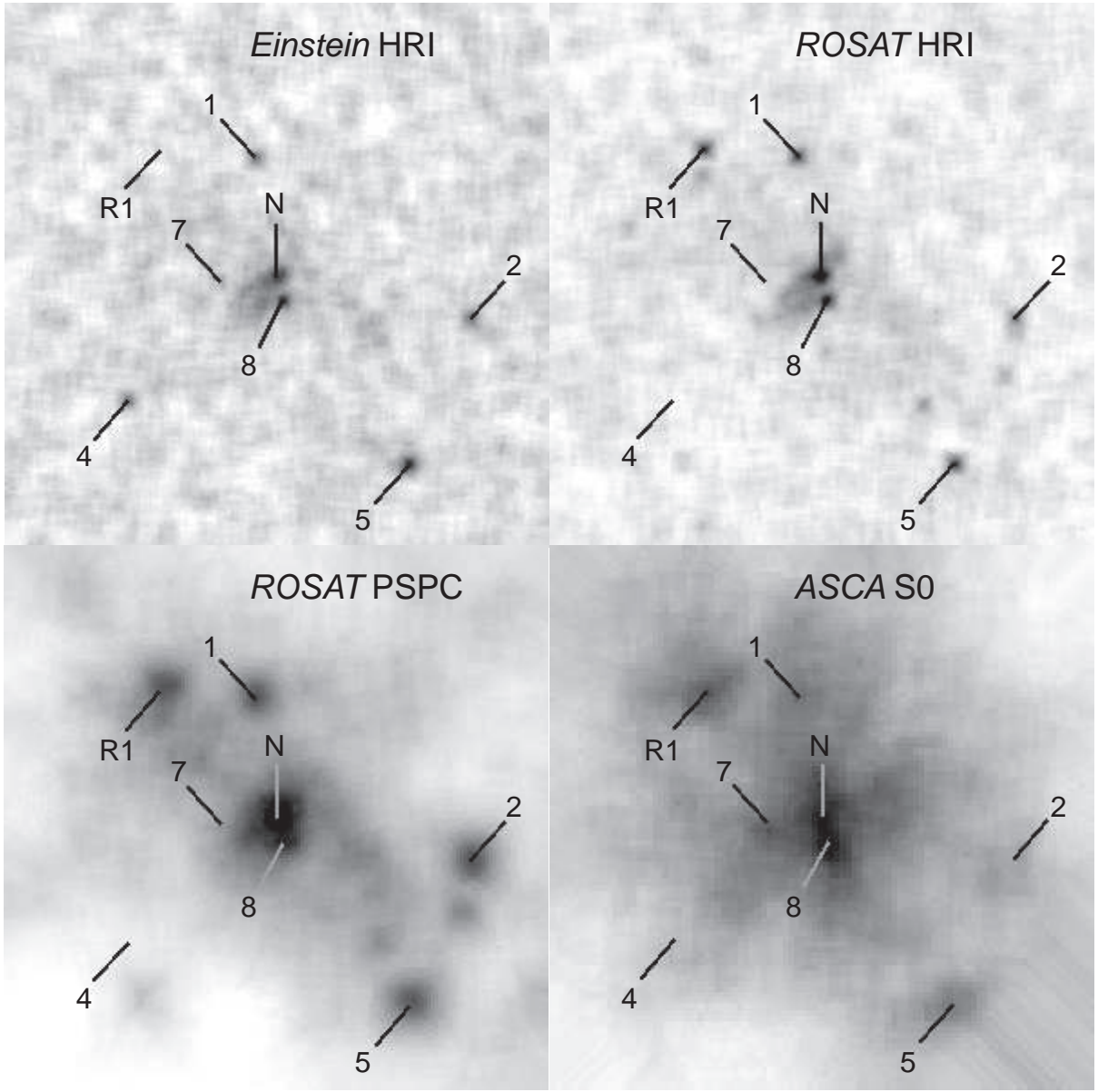}{6in}{0}{100}{100}{-312}{-200}
\caption{{\it Einstein} HRI (0.2-4.0 keV, 7/5-9/79; top left), {\it ROSAT} 
HRI (0.1-2.2
keV; 6/7/92; top right), {\it ROSAT} PSPC (0.5-2.0 keV; 12/6/91-5/6/92; bottom
left), and {\it ASCA} S0 (2-10 keV; 5/25/93; bottom right) images of NGC 253.
Each image is 10' by 10' and has been adaptively smoothed to produce a SNR of
$\gtrsim 5$ at each pixel (without background subtraction). The sources 
discussed in %\markcite{a10}
Fabbiano \& Trinchieri (1984) are marked (sources 3
and 6 lie outside of the FOV of these images) along with the X-ray nucleus
(`N') and the ``new'' {\it ROSAT} source (`R1').}
\end{figure*}

The authors wish to thank all of the members of {\it ASCA} team who have made
this work possible. AP wishes to acknowledge the support of the NASA Graduate
Student Research Program.
This research has made use of the NASA's Astrophysics Data 
System Abstract Service and the NASA/IPCA Extragalactic 
Database (NED) which is operated by the Jet Propulsion Laboratory, Caltech, 
under contract with NASA. This
work will be included in a dissertation to be submitted to the Graduate School,
University of Maryland, in partial fulfillment of the requirements for the
Ph.D. degree in Physics.

\appendix
\section*{Appendix: Computation of the {\it ASCA} X-Ray Telescope (XRT) 
Response for Extended Sources}
As stated in the text, the X-ray emission in both M82 and NGC 253 is extended
while the telescope response was computed assuming it is point-like.  In this
appendix we show that properly taking the extended nature of the sources
into account does not significantly impact the results of this paper.  In
X-ray spectral fitting, the XRT response is taken into account by the
creation of a table of effective area as a function of energy, which
is then used to convert the observed counting rate in each spectral bin
to a flux.  This table, hereafter referred to as an ARF (ancillary response
file), also applies a correction for the counts 
scattered out of the source region (6' circular regions were used for both the
SIS and GIS in this paper).

\subsection*{The {\it ASCA} XRT Response for a Point Source}
First, we describe the ``standard'' procedure used for the creation of an
ARF appropriate for a point source. The telescope
point-spread function (PSF) is summed over
the source region to determine percentage of source photons contained in the
source region.  The PSF sum is then multiplied by the telescope effective
area.  The PSF and effective area are functions of both energy and the off-axis
angle of the source.  Both the PSF and effective area are interpolated from
values contained in calibration look-up tables, computed using a Monte-Carlo
procedure based on a raytracing program.  Although the XRT PSF has a
cross-like shape, the calibration tables contain azimuthally-averaged
values, an approach that is usually a good
approximation as long as most of the source
counts lie in a symmetric (and preferably circular) region. The mirrors contain
a heat shield, made up of a thin plastic film, to protect against temperature
gradients, and the computed effective area is reduced to take photon losses
through the 
shield into account. The gas cell of the GIS contains a Beryllium window
held in place by a supporting grid, and transmission through the grid
and Be must also be taken into account.  Note that the thicknesses of both
the grid
and Be vary across the GIS detectors. Also, unlike the SIS, the GIS introduce
a broadening of the PSF.  The instrumental PSF of the GIS is approximately a
Gaussian whose FWHM $\sim 0.5\sqrt{\frac{5.9 \rm \ keV}{E}}$ arcminutes, with
the addition of a ``tail'' at energies $\gtrsim 4$ keV.  Accordingly,
the XRT PSF is modified by the grid and Be transmission and then convolved
with the GIS PSF, which is modeled by the sum of two-dimensional Gaussians.
Hereafter, we refer to ARFs computed with this approach as standard ARFs.

\subsection*{Computation of the {\it ASCA} Response to an Extended Source}
There are two ways in which standard ARFs may be inappropriate for an
extended source.  First, more source counts would be scattered out of the
source region than is the case for a point source.  Not taking this into
account would result in an underestimate of the source flux.  Second, by
definition, an extended source subtends a distribution of off-axis angles, so
determining the XRT effective area and PSF for a single off-axis angle may
be incorrect.  Because both the effective area and PSF have an energy
dependence, the spectral results may be affected as well as the flux
estimates.  Because the X-ray emission from M82 and NGC 253 is relatively
compact and weak, it is
likely that statistical errors will dominate that the effects described above,
with the possible exception that the fluxes may be
underestimated. To check this in detail, we adopted a
Monte-Carlo approach.  As is the case in the point-source procedure, we
divided the
bandpasses of the SIS and GIS into the same number of energy bins as
is used in the spectral fitting, namely 1180 in the case of the SIS
and 201 in the case of the GIS.  For each bin at least 10000 photons were
raytraced through the mirror (the raytracing code was developed by Peter
Serlemitsos during the design of the {\it ASCA} mirrors, and, when
azimuthally-averaged, is in good agreement with the look-up tables described
above and the radial profile of an {\it ASCA} PV phase observation of
3C 273).  The position (on the sky) of each photon was chosen at random from
{\it ROSAT} PSPC images of M82 and NGC 253, weighted by the intensity at
each pixel.  This approach results in photons incident on the XRT with
a distribution in off-axis angles consistent with those expected from
the PSPC images.  A 12' by 12' image was used for M82 and a 10' by 10'
image was used for NGC 253.  The bandpass of each image was limited to
$\sim 0.5-2.0$ keV, and the background was {\it not} subtracted, in
order to conservatively overestimate the source extent (note that the
source extent is also overestimated since the {\it ROSAT} PSPC PSF is
not removed with this procedure, although the {\it ROSAT} PSF FWHM is
$\sim 10''$, somewhat smaller than the {\it ASCA} XRT PSF FWHM of
$\sim 50''$).

In the case of the GIS, the instrumental PSF was introduced
by adding ``noise'' (i.e., Gaussian random deviates) to the raytraced photon
positions.  The standard
deviations of the deviates were computed with the same
algorithm used in the standard ARF procedure.  Also in the case of the
GIS, photons were
rejected to account for the grid and Be transmission (i.e., a photon
was rejected if a uniform deviate exceeded the product of the Be and
grid transmission at the photon's position).  The percentage
of photons ``detected'' in the source region, reduced by the thermal
shield transmission (as is done in the point source case), was then
multiplied by the geometrical area of the mirror to derive an effective area.
Hereafter, we refer to ARFs computed with this procedure as ``raytraced''
(although, as stated above, the ``standard'' ARF procedure also
{\it indirectly} involves raytracing, in the computation of the look-up
tables).

\begin{figure}
%\plotfiddle{m82_n253_paper_fig6.ps}{4in}{0}{100}{100}{0}{0}
\plotfiddle{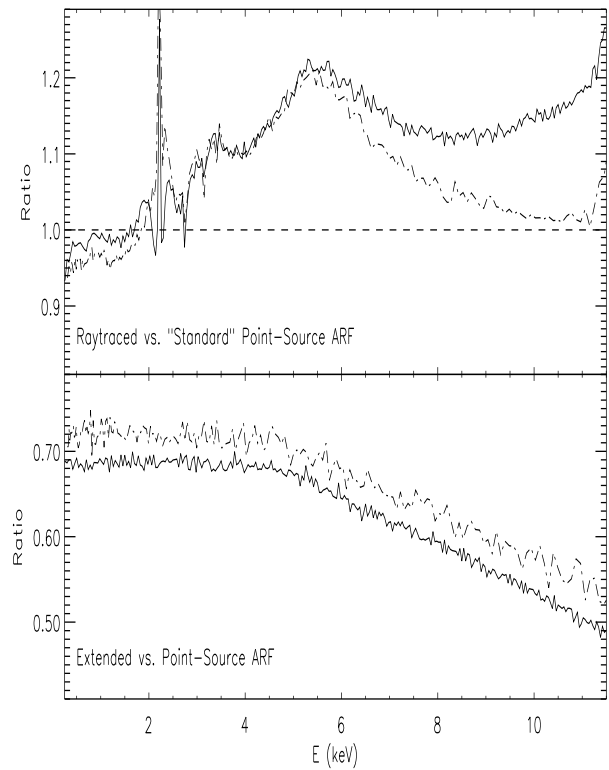}{4in}{0}{45}{85}{-143}{-310}
\caption{({\it top}) The ratio of 
ARFs for a point source computed using the raytracing algorithm described in the
text and the ``standard'' procedure using look-up tables for the XRT effective
area and PSF.  The ratio for ARFs appropriate for the S0 (solid line) and 
G2 (dashed-line) M82 data are shown.
The largest source of discrepancy between the raytraced and ``standard''
ARFs arises from differences in
the raytracing code used in this analysis and the raytracing used to produce 
the look-up tables. ({\it bottom}) The ratio of the ARFs computed assuming 
that M82
is extended and assuming that M82 is a point source in the case of S0 
(solid-line) and S2 (dashed line).  Here, both the point-source and
extended-source ARFs were raytraced to remove the discrepancies
between the raytraced and ``standard'' procedures (shown in the top panel).}
\end{figure}

Initially, ARFs were raytraced assuming that M82 and NGC 253 are point sources
for comparison with the standard ARFs used in the fits given in the text.
The ratio of the raytraced ARFs to the standards ARFs, shown in the top
panel of figure 6 for the case of the M82 S0 and S2 ARFs, gives the
systematic
differences between the two procedures described above.
The largest source of discrepancy between the two methods is due to differences
in the raytracing code used here and the raytracing used to produce the
look-up tables, particularly in the choice of optical constants for gold
(the reflecting material used in the XRT).  Other sources of discrepancy are
inherent to the procedures employed, i.e., the XRT PSF is not
azimuthally-averaged in the raytraced case.  The top panel of figure 6 shows
that, with the exception of a sharp peak around the Au M edge at 2.2 keV, the
two procedures are in agreement below 4 keV to within 10\%. Above 4 keV there
is a broad ``hump'' in the ratio plot which peaks at a level of $\sim 20\%$.
These are issues which are currently being resolved by the
instrument teams and will be reported in a future issue of {\it ASCANews}.
For comparison, the
statistical errors in the ASCA data below 4 keV are typically on the order
of 6\% in the case of M82 and 10\% in the case of NGC 253.  As is evident in
Figure 2, the statistical errors become somewhat larger above 4 keV since
both the XRT and instrumental responses and the source fluxes drop off rapidly
above 4 keV, particularly in the case of NGC 253 which has a steeper spectrum
than M82.

%It has
%recently become evident that some of the {\it ASCA} calibration incorporated
%in the instrumental responses may in fact be corrections to uncertainty
%in the XRT response.  Because of this, the instrumental response files are
%only appropriate for ARFs created using the look-up tables described above.
%To correct for this, the above procedure was repeated
%assuming the M82 and NGC 253 are point sources (i.e., all of the photons
%were raytraced assuming a single off-axis angle).  Each extended source 
%ARF was then multiplied by the ratio of the point-source ARF created
%using the ``standard'' procedure and the raytraced point-source ARF.  This
%effectively ``renormalizes'' the extended source ARFs to the XRT response
%appropriate for the instrumental responses.
%
\subsection*{Implications for M82 and NGC 253}
The bottom panel of figure 6 shows that ratio of ARFs produced with the
raytracing procedure for an extended source (with photons selected from the
PSPC image, as described above) and a point-source (raytraced ARFs were
computed in both cases to remove the discrepancy between the raytraced and
standard ARFs) in the case of the M82 SIS0 (solid line) and GIS2 (dashed line)
ARFs.  Note that the extended nature of the source only impacts the overall
normalization below $\sim 4$ keV, and results in a gradual drop off in net
effective area above 4 keV (a result of the larger energy dependence of the
XRT PSF and effective area above 4 keV).  Note that this drop in effective area
partially offsets the ``hump'' above 4 keV noted above.  Indeed, in the case
of NGC 253, repeating the four-instrument Raymond-Smith plus power-law fit
with the extended source ARFs results in best-fitting parameters that are
essentially identical to those discussed in the text where the standard ARFs
were used (i.e., the statistical errors dominate any systematic errors).  When
the double Raymond-Smith fit to the M82 data is repeated with the
extended-source
ARFs, the soft component parameters do not change significantly, although the
inferred 0.5-2.0 keV flux is increased by $\sim 40\%$.  Because our procedure 
deliberately overestimates the source extent, this is a conservative upper
limit to the 0.5-2.0 keV flux from M82.  On the other hand, the hard
component spectral parameters are modified slightly when extended-source ARFs
are used, with the best-fitting $N_H$ decreasing by $\sim 5 \times 10^{21} \
\rm cm^{-2}$, the temperature increasing by $\sim 3$ keV and
the abundance increasing from $\sim 0.2$ solar to $\sim 0.3$
solar.  Nevertheless, these parameters are still well within the statistical
errors quoted in the text.  As with the soft component, the extended-source
fits suggest that the 2-10 keV flux, dominated by the hard component, may be
underestimated by $\sim 27\%$.  However, Tsuru {\it et al.} (1997) report that
the flux 
associated with the hard component in M82 is not resolved by {\it ASCA}.  
Recomputing the extended-source ARFs by selecting the soft-component photons
from
the PSPC image and assuming that the hard component is point-like eliminates
the changes in the hard-component fit parameters, and results in a
2-10 keV flux $\sim 7\%$ {\it lower} than that inferred from the standard ARF
fits (this is not surprising since, as shown in the top panel of Figure
6, the raytraced point-source ARF gives a larger effective area at high 
energies than the standard ARFs).
Thus, the only parameter significantly affected by properly computing the
response of {\it ASCA} to an extended source is the 0.5-2.0 keV flux inferred
for M82, which does not impact any of the conclusions of this paper.

\newpage

\newpage
%\begin{figure}
%\plotone{m82_n253_letter_fig2.eps}
%\caption{}
%\end{figure}
%\begin{figure}
%\plotone{m82_n253_letter_fig3.eps}
%\caption{}
%\end{figure}


\begin{references} 
\reference{a1} Bhattacharya, D., {\it et al.} 1994, \apj, 437, 173 
\reference{a2} Bregman, J., Schulman, E., \& Tomisaka, K. 1995, \apj, 439, 155 
\reference{a3} Calzetti, D., Kinney, A., \& Storchi-Bergmann, T. 1996, \apj,
458, 132
\reference{a4} Carral, P., Hollenbach, D., Lord, S., Colgan, S., Haas, M., 
Rubin, R., \& Erickson, E. 1994, \apj, 423, 223
\reference{a102} Carilli, C., Holdaway, M., Ho, P., \& De Pree, C. 1992, \apj,
399, L59
\reference{a5} Collura, A., Reale, F., Schulman, E., \& Bregman, J. 1994, \apj,
420, L63 
\reference{a6} Cowie, L. \& Songaila, A. 1986, \araa, 24, 499
\reference{a7} Dahlem, M., Heckman, T., \& Weaver, K. 1997, \apj, in press
\reference{a100} Dahlem, M., Heckman, T., \& Fabbiano, G. 1995, \apj, 442, L49
\reference{a8} Davidge, T. \& Pritchet, C. 1990, AJ, 100, 102
\reference{a9} Fabbiano, G. 1988, \apj, 330, 672
\reference{a10} Fabbiano, G. \& Trinchieri, G. 1984, \apj, 286, 491
\reference{a11} Freedman, W. {\it et al.} 1994, \apj, 427, 628
\reference{a12} Hamilton, A., Sarazin, C., \& Chevalier, R. 1983, \apjs, 51, 
115
\reference{a13} Heckman, T., Armus, L., \& Miley, G. 1990, \apjs, 74, 833
\reference{a14} Huang, Z., Thuan, T., Chevalier, R., Condon, J., \& Yin, Q. 
1994, \apj, 424, 114
\reference{a15} Lord, S., Hollenbach, D., Haas, M., Rubin, R., Colgan, S., \& 
Erickson, E. 1996, \apj, 465, 703
\reference{a16} Mewe, R., Gronenschild, E., \& Van Den Oord, G. 1985, \aaps, 
62, 197
\reference{a17} Moran, E. \& Lehnert, M. 1997, \apj, in press
\reference{a18} Nakai, N. {\it et al.} 1987, \pasj, 39, 685
\reference{a19} Ohashi, T., Makishima, K., Tsuru, T., Takano, S., Koyama, K, 
\& Stewart, G. 1990, \apj, 365, 180
\reference{a20} Paglione, T., Tosaki, T., \& Jackson, J. 1996, \apj, 454, L117
\reference{a21} Petre, R. 1993, in The Nearest Active Galaxies, eds. J. 
Beckman, L. Colina, and H. Netzer, 117
\reference{a22} Rieke, G., Lebofsky, M., Thompson, R., Low, F., \& Schneider,
D. 1980, \apj, 238, 24
\reference{a23} Schaaf, R., Pietsch, W., Biermann, P., Kronberg P., \& 
Schmutzler, T. 1989, \apj, 336, 722
\reference{a24} Seaquist, E. \& Odegard, N. 1991, \apj, 369, 320
\reference{a25} Serlemitsos, P., Ptak, A., \& Yaqoob, T. 1996, in The Physics 
of LINERs in View of Recent Observations, eds., M. Eracleous, A. Koratkar, C.
Leitherer, and L. Ho, 70
\reference{a26} Stark, A. A., Gammie, C. F., Wilson, R. W., Bally, J., Linke, 
R., Heiles, C., \& Hurwitz,M. 1992, \apjs, 79, 77
\reference{a27} Strickland, D., Ponman, T., \& Stevens, I. 1997, \aa, in press
\reference{a28} Suchkov, A., Berman, V., Heckman, T., \& Balsara, D. 1996, 
\apj, 463, 528
\reference{a101} Yaqoob, T., Serlemitsos, P., Ptak, A., Mushotzky, R., 
Kunieda, H. \& Terashima, Y. 1995, ApJ, 455, 508
\reference{a29} Tanaka, Y. 1989 in 23rd ESLAB Symp. on Two Topics in X-ray 
Astronomy, eds. J. Hunt and B. Battrick, 3
\reference{a30} Tanaka, Y., Inoue, H., \& S. Holt 1994, \pasj, 46, L37
\reference{a31} Telesco, C. 1988, \araa, 26, 343
\reference{a32} Thielemann, F., Nomoto, K., \& Hashimoto, M. 1996, \apj, 460, 
408
\reference{a33} Tomisaka, K. \& Bregman, J. 1993, \pasj, 45, 513
%\reference{a34} Tsuru,T., {\it et al.} 1997a in The {\it ASCA} 3rd Anniv. 
%Symp., in press
\reference{a35} Tsuru, T., {\it et al.} 1997, \pasj, submitted
\reference{a36} Tsuru, T., Ohashi, T., Makishima, K., Mihara, T., \& Kondo, H. 
1990, \pasj, 42, L75
\reference{a37} Watson, M., Stanger, V., \& Griffiths, R. 1984, \apj, 286, 144
\end{references}
\end{document}